\documentclass[aps,prl,twocolumn,lineno,groupedaddress]{revtex4} 
\usepackage{graphicx}  
\usepackage{dcolumn}   
\usepackage{bm}        
\usepackage{amssymb}   

\begin{document}

\hspace{5.2in} \mbox{FERMILAB-PUB-06-109-E}
  
\title{Search for particles decaying into a $Z$ boson and a photon in $p\overline{p}$ collisions at $\sqrt{s} = 1.96$ TeV }

%
\author{                                                                      
V.M.~Abazov,$^{36}$                                                           
B.~Abbott,$^{76}$                                                             
M.~Abolins,$^{66}$                                                            
B.S.~Acharya,$^{29}$                                                          
M.~Adams,$^{52}$                                                              
T.~Adams,$^{50}$                                                              
M.~Agelou,$^{18}$                                                             
J.-L.~Agram,$^{19}$                                                           
S.H.~Ahn,$^{31}$                                                              
M.~Ahsan,$^{60}$
% ...
\\                                                                            
\vskip 0.30cm                                                                 
\centerline{(D\O\ Collaboration)}                                             
\vskip 0.30cm                                                                 
}                                                                             
\affiliation{                                                                 
\centerline{$^{1}$Universidad de Buenos Aires, Buenos Aires, Argentina}       
\centerline{$^{2}$LAFEX, Centro Brasileiro de Pesquisas F{\'\i}sicas,         
                  Rio de Janeiro, Brazil}                                     
\centerline{$^{3}$Universidade do Estado do Rio de Janeiro,                   
                  Rio de Janeiro, Brazil}                                     
\centerline{$^{4}$Instituto de F\'{\i}sica Te\'orica, Universidade            
                  Estadual Paulista, S\~ao Paulo, Brazil}                     
\centerline{$^{5}$University of Alberta, Edmonton, Alberta, Canada,           
                  Simon Fraser University, Burnaby, British Columbia, Canada,}
\centerline{York University, Toronto, Ontario, Canada, and                    
                  McGill University, Montreal, Quebec, Canada}                
\centerline{$^{6}$Institute of High Energy Physics, Beijing,                  
                  People's Republic of China}                                 
\centerline{$^{7}$University of Science and Technology of China, Hefei,       
                  People's Republic of China}                                 
\centerline{$^{8}$Universidad de los Andes, Bogot\'{a}, Colombia}             
\centerline{$^{9}$Center for Particle Physics, Charles University,            
                  Prague, Czech Republic}                                     
\centerline{$^{10}$Czech Technical University, Prague, Czech Republic}        
\centerline{$^{11}$Center for Particle Physics, Institute of Physics,         
                   Academy of Sciences of the Czech Republic,                 
                   Prague, Czech Republic}                                    
\centerline{$^{12}$Universidad San Francisco de Quito, Quito, Ecuador}        
\centerline{$^{13}$Laboratoire de Physique Corpusculaire, IN2P3-CNRS,         
                   Universit\'e Blaise Pascal, Clermont-Ferrand, France}      
\centerline{$^{14}$Laboratoire de Physique Subatomique et de Cosmologie,      
                   IN2P3-CNRS, Universite de Grenoble 1, Grenoble, France}    
\centerline{$^{15}$CPPM, IN2P3-CNRS, Universit\'e de la M\'editerran\'ee,     
                   Marseille, France}                                         
\centerline{$^{16}$IN2P3-CNRS, Laboratoire de l'Acc\'el\'erateur              
                   Lin\'eaire, Orsay, France}                                 
\centerline{$^{17}$LPNHE, IN2P3-CNRS, Universit\'es Paris VI and VII,         
                   Paris, France}                                             
\centerline{$^{18}$DAPNIA/Service de Physique des Particules, CEA, Saclay,    
                   France}                                                    
\centerline{$^{19}$IReS, IN2P3-CNRS, Universit\'e Louis Pasteur, Strasbourg,  
                    France, and Universit\'e de Haute Alsace,                 
                    Mulhouse, France}                                         
\centerline{$^{20}$Institut de Physique Nucl\'eaire de Lyon, IN2P3-CNRS,      
                   Universit\'e Claude Bernard, Villeurbanne, France}         
\centerline{$^{21}$III. Physikalisches Institut A, RWTH Aachen,               
                   Aachen, Germany}                                           
\centerline{$^{22}$Physikalisches Institut, Universit{\"a}t Bonn,             
                   Bonn, Germany}                                             
\centerline{$^{23}$Physikalisches Institut, Universit{\"a}t Freiburg,         
                   Freiburg, Germany}                                         
\centerline{$^{24}$Institut f{\"u}r Physik, Universit{\"a}t Mainz,            
                   Mainz, Germany}                                            
\centerline{$^{25}$Ludwig-Maximilians-Universit{\"a}t M{\"u}nchen,            
                   M{\"u}nchen, Germany}                                      
\centerline{$^{26}$Fachbereich Physik, University of Wuppertal,               
                   Wuppertal, Germany}                                        
\centerline{$^{27}$Panjab University, Chandigarh, India}                      
\centerline{$^{28}$Delhi University, Delhi, India}                            
\centerline{$^{29}$Tata Institute of Fundamental Research, Mumbai, India}     
\centerline{$^{30}$University College Dublin, Dublin, Ireland}                
\centerline{$^{31}$Korea Detector Laboratory, Korea University,               
                   Seoul, Korea}                                              
\centerline{$^{32}$SungKyunKwan University, Suwon, Korea}                     
\centerline{$^{33}$CINVESTAV, Mexico City, Mexico}                            
\centerline{$^{34}$FOM-Institute NIKHEF and University of                     
                   Amsterdam/NIKHEF, Amsterdam, The Netherlands}              
\centerline{$^{35}$Radboud University Nijmegen/NIKHEF, Nijmegen, The          
                  Netherlands}                                                
\centerline{$^{36}$Joint Institute for Nuclear Research, Dubna, Russia}       
\centerline{$^{37}$Institute for Theoretical and Experimental Physics,        
                   Moscow, Russia}                                            
\centerline{$^{38}$Moscow State University, Moscow, Russia}                   
\centerline{$^{39}$Institute for High Energy Physics, Protvino, Russia}       
\centerline{$^{40}$Petersburg Nuclear Physics Institute,                      
                   St. Petersburg, Russia}                                    
\centerline{$^{41}$Lund University, Lund, Sweden, Royal Institute of          
                   Technology and Stockholm University, Stockholm,            
                   Sweden, and}                                               
\centerline{Uppsala University, Uppsala, Sweden}                              
\centerline{$^{42}$Physik Institut der Universit{\"a}t Z{\"u}rich,            
                   Z{\"u}rich, Switzerland}                                   
\centerline{$^{43}$Lancaster University, Lancaster, United Kingdom}           
\centerline{$^{44}$Imperial College, London, United Kingdom}                  
\centerline{$^{45}$University of Manchester, Manchester, United Kingdom}      
\centerline{$^{46}$University of Arizona, Tucson, Arizona 85721, USA}         
\centerline{$^{47}$Lawrence Berkeley National Laboratory and University of    
                   California, Berkeley, California 94720, USA}               
\centerline{$^{48}$California State University, Fresno, California 93740, USA}
\centerline{$^{49}$University of California, Riverside, California 92521, USA}
\centerline{$^{50}$Florida State University, Tallahassee, Florida 32306, USA} 
\centerline{$^{51}$Fermi National Accelerator Laboratory,                     
            Batavia, Illinois 60510, USA}                                     
\centerline{$^{52}$University of Illinois at Chicago,                         
            Chicago, Illinois 60607, USA}                                     
\centerline{$^{53}$Northern Illinois University, DeKalb, Illinois 60115, USA} 
\centerline{$^{54}$Northwestern University, Evanston, Illinois 60208, USA}    
\centerline{$^{55}$Indiana University, Bloomington, Indiana 47405, USA}       
\centerline{$^{56}$University of Notre Dame, Notre Dame, Indiana 46556, USA}  
\centerline{$^{57}$Purdue University Calumet, Hammond, Indiana 46323, USA}    
\centerline{$^{58}$Iowa State University, Ames, Iowa 50011, USA}              
\centerline{$^{59}$University of Kansas, Lawrence, Kansas 66045, USA}         
\centerline{$^{60}$Kansas State University, Manhattan, Kansas 66506, USA}     
\centerline{$^{61}$Louisiana Tech University, Ruston, Louisiana 71272, USA}   
\centerline{$^{62}$University of Maryland, College Park, Maryland 20742, USA} 
\centerline{$^{63}$Boston University, Boston, Massachusetts 02215, USA}       
\centerline{$^{64}$Northeastern University, Boston, Massachusetts 02115, USA} 
\centerline{$^{65}$University of Michigan, Ann Arbor, Michigan 48109, USA}    
\centerline{$^{66}$Michigan State University,                                 
            East Lansing, Michigan 48824, USA}                                
\centerline{$^{67}$University of Mississippi,                                 
            University, Mississippi 38677, USA}                               
\centerline{$^{68}$University of Nebraska, Lincoln, Nebraska 68588, USA}      
\centerline{$^{69}$Princeton University, Princeton, New Jersey 08544, USA}    
\centerline{$^{70}$State University of New York, Buffalo, New York 14260, USA}
\centerline{$^{71}$Columbia University, New York, New York 10027, USA}        
\centerline{$^{72}$University of Rochester, Rochester, New York 14627, USA}   
\centerline{$^{73}$State University of New York,                              
            Stony Brook, New York 11794, USA}                                 
\centerline{$^{74}$Brookhaven National Laboratory, Upton, New York 11973, USA}
\centerline{$^{75}$Langston University, Langston, Oklahoma 73050, USA}        
\centerline{$^{76}$University of Oklahoma, Norman, Oklahoma 73019, USA}       
\centerline{$^{77}$Oklahoma State University, Stillwater, Oklahoma 74078, USA}
\centerline{$^{78}$Brown University, Providence, Rhode Island 02912, USA}     
\centerline{$^{79}$University of Texas, Arlington, Texas 76019, USA}          
\centerline{$^{80}$Southern Methodist University, Dallas, Texas 75275, USA}   
\centerline{$^{81}$Rice University, Houston, Texas 77005, USA}                
\centerline{$^{82}$University of Virginia, Charlottesville,                   
            Virginia 22901, USA}                                              
\centerline{$^{83}$University of Washington, Seattle, Washington 98195, USA}  
}                                                                             

\date{May 15, 2006}

\begin{abstract}
We present the results of a search for a new particle $X$ produced in 
$p\bar{p}$ collisions at $\sqrt{s}=1.96$ TeV and subsequently decaying to 
$Z\gamma$.  The search uses 0.3 fb$^{-1}$ of data collected with the D\O\ 
detector at the Fermilab Tevatron Collider.  We set limits on the production cross 
section times the branching fraction 
$\sigma(p \bar{p} \rightarrow X) \times B(X\rightarrow Z\gamma)$ 
that range from 0.4 to 3.5 pb at the 95\% C.L.
for $X$ with invariant masses between 100 and 1000 GeV/$c^2$, over a wide 
range of $X$ decay widths.
\end{abstract}

\maketitle


There is considerable evidence that the
standard model (SM) is incomplete~\cite{Quigg:2005hr}.  Signs of new 
physics may appear in
the form of a new particle ($X$).  If $X$ is a scalar,  
pseudo-scalar, or tensor, its decay to lepton pairs might be highly 
suppressed, but it could have a large decay branching fraction ($B$) to 
the di-boson final state $Z\gamma$.
A search for $X$ in the $Z\gamma$ final state thus complements previous 
searches (for example~\cite{Feldman:2006ce}) for production of a new 
vector boson in the lepton pair decay mode.

Events with pairs of vector bosons have been studied as tests of 
the SM of electroweak interactions. 
Specifically, the $Z$ plus photon final state ($Z\gamma$) has been investigated 
by the D\O\ ~\cite{us,otherD0} and CDF~\cite{cdf} collaborations using
$p\bar{p}$ collisions and by the LEP collaborations~\cite{delphi,L3,
Abbiendi:2000cu} using
$e^+e^-$ collisions. In these cases, the measured cross section and photon
energy distribution were used to
set limits on anomalous $Z$-photon couplings, but no explicit searches
for new particles decaying to $Z\gamma$ were performed. The L3 
   Collaboration ~\cite{L3higgs} searched for 
   Higgs boson production, with subsequent decay of the Higgs 
   to $Z\gamma$, in electron-positron collisions at the LEP2 collider, and 
   set cross section limits for Higgs boson masses up to 190 GeV$/c^2$.

\begin{figure}
\includegraphics[scale=0.45]{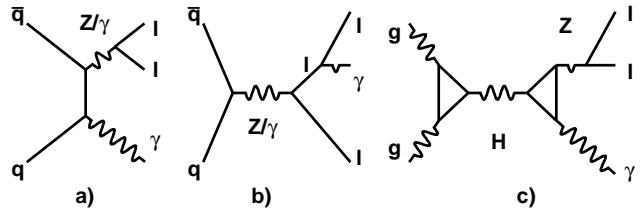}
\caption{
The Feynman diagrams for standard model sources of  dilepton plus 
$\gamma$ events are shown. Diagram a) shows $q\bar{q} \rightarrow Z$-boson
plus $\gamma$, where the photon is radiated from the quark or anti-quark. 
Diagram b) shows $q\bar{q} \rightarrow Z/\gamma$, where the photon is radiated
from one of the $Z$ boson's decay products. Diagram c) shows Higgs 
production and decay into a $Z$ boson and a photon.
\label{fig:feyn}}
\end{figure}

In the SM, the dilepton plus $\gamma$ final state, including $Z\gamma$, is 
expected to be produced through
radiative processes (Figs.~\ref{fig:feyn}a and b).  
In addition, this final state is also 
expected from Higgs boson production and decay (Fig.~\ref{fig:feyn}c).  
Although the Higgs boson mass is unknown and the predicted 
$H\to Z\gamma$ 
branching fraction is ${\cal O}(10^{-3})$, extensions to the 
SM can significantly increase this branching fraction~\cite{spira,Buscher:2005re,Akeroyd:2005pr,Arik:2005ed}.  
Other SM extensions predict new particles that decay 
into $Z\gamma$.  
For example, a $Z^\prime$ boson can
decay radiatively to a $Z$ boson and a photon~\cite{Kozlov:2005rj}.  In models
with a fourth generation of fermions, a top and anti-top quark bound state (toponium) may
exist~\cite{Ono:1983tf,Cakir:2004nh}, and this state can decay to $Z\gamma$. 
In theories
with compact extra dimensions, massive Kaluza-Klein spin-2 gravitons can
also decay to the $Z\gamma$ final state~\cite{Davoudiasl:2000wi}. 
The presence of resonance behavior in the $Z\gamma$ final state can thus signal the presence of a wide variety of new physics.  In order to make quantitative statements, we will assume that this new physics manifests itself in the form of a spin 0 particle.

A large sample of $Z\gamma$ events 
has been collected by the D\O\ experiment and analyzed to measure
the $Z\gamma$ cross section and set limits on anomalous 
$ZZ\gamma$ and $Z\gamma\gamma$ couplings~\cite{us}.
The Fermilab Tevatron Collider provides a higher energy reach than that 
available to  
previous experiments, and so this sample deserves further scrutiny. 
Experimentally, $Z$ bosons are identified through their decay to
charged lepton pairs ($\ell\ell=ee$ or $\mu\mu$).  Photons are 
measured with high
precision from their electromagnetic showers.  The
$Z\gamma$ final state has small backgrounds.  We focus on, by 
tightening kinematic selection criteria, and study the mass distribution of the 
$\ell\ell\gamma$ events in a sample of 0.3 fb$^{-1}$ of 
$p\overline{p}$ collision data collected with the D\O\ Run II detector 
from April 2002 to June 2004 at the Fermilab Tevatron Collider 
at $\sqrt{s}$ = 1.96 TeV.   

The D\O\ detector~\cite{run2det} includes a central tracking system, composed 
of a 
silicon microstrip tracker and a central fiber tracker, both 
located within a 2~T superconducting solenoidal magnet and optimized for tracking and vertexing capability at 
pseudorapidities \cite{coord} of $|\eta|<2.5$. Three 
liquid argon and uranium calorimeters provide coverage up to 
$|\eta|\approx 4.2$: a central section, and two end calorimeters. 
A muon system resides beyond 
the calorimetry, and consists of tracking detectors,
scintillation counters, and a 1.8~T toroid with coverage for $|\eta|<2$.
Luminosity is measured using 
scintillator arrays located in front of the end calorimeter cryostats, covering
$2.7 < |\eta| < 4.4$.
Trigger and data acquisition systems are designed to accommodate the high
luminosities of the Run II Tevatron. 

The analysis is conducted in two channels, one where the $Z$ boson decays into 
electrons and the other where it decays into muons.
Electron candidate events are required to 
satisfy one of a series of single electron triggers.
The electron channel requires that electron candidates 
be isolated in the calorimeter, have 
longitudinal and transverse energy deposition profiles consistent with those of an electron, have
a transverse momentum $p_{T} >$ 15 GeV/$c$, and 
be contained in either the central calorimeter (CC, $|\eta|$ $<$ 1.1) 
or one of the end calorimeters (EC, 1.5 $<$ $|\eta|$ $<$ 2.5) and not in 
the transition region between the central and the end calorimeters.  
If an electron candidate is in the CC, it is required to have a 
spatially matched track from the central tracker. 
One of the electrons must have $p_{T} >$ 25 GeV/$c$.  
The efficiency for a di-electron candidate to satisfy the trigger and for both
electrons to satisfy all quality requirements lead to an event efficiency 
of 0.69 $\pm$ 0.05 if both electrons are in the CC and 0.78 $\pm$ 0.05 
if one electron is in the EC. 
Events with both electron candidates in the EC are not considered due to a small expected number of events from X and large backgrounds.  
These efficiencies are measured with the inclusive $Z$ boson candidate events.

Muon candidate events must pass one of a suite of 
single or di-muon triggers.  The muon channel requires two candidate muons with $p_{T} > $15 GeV/$c$ and opposite
charge.  Both muons must be matched to tracks found in the central tracker. 
The background from heavy flavor production is suppressed by requiring
the muon candidates to be isolated.  The background from cosmic rays is
suppressed by requiring that the muons come from the interaction region and
are not exactly back-to-back.
The efficiency for di-muon event selection
and trigger is 0.84 $\pm$ 0.05 per event.  This efficiency is
measured with $Z$ boson candidate events.

Photon candidates must be isolated in the calorimeter and
tracker, have longitudinal and transverse shapes in the calorimeter
consistent with 
those of a photon, have $p_{T} >$ 25 GeV/$c$, and 
be contained in the central calorimeter ($|\eta|$ $<$ 1.1). 
The efficiency is around 0.85 at 25 GeV/$c$ and rises to a plateau of
0.90 at 35 GeV/$c$.

Both di-electron and di-muon candidate events are further required to have a di-lepton mass greater than 75 GeV/$c^2$,
and a photon separated from both leptons by 
$\Delta \cal{R} >$ 0.7 ~\cite{deltaR}.  These requirements reduce the 
contribution
from events in which a final state lepton radiates a photon.
The detector acceptance times particle identification
and trigger efficiency, for all requirements described, rises from 
about 18\% to about 20\% for masses from 100 to 800 GeV/$c^2$ and at 
higher masses decrease.  
At mass greater than 800 GeV/c$^2$, a significant number of leptons fall
within the isolation region of the other lepton, and charge misidentification
becomes significant.  The uncertainty on these are
the dominant contributors to the systematic uncertainty on the expected 
number of signal candidates. At 800 GeV/c$^2$, the uncertainty
is approximately $10\%$ and at 1000 GeV/c$^2$, the uncertainty has 
risen to $40\%$. 

To improve the di-lepton-photon mass resolution in the muon  
channel,  the muon transverse momenta are adjusted by employing a 
one-constraint kinematic fit that forces the di-muon mass to equal the 
on-shell $Z$-boson mass.  This constraint is only enforced if the 
fit has $\chi^2/$d.o.f.$ < 7 $. 
Monte Carlo studies show this technique improves the 
three-body mass resolution from 6.7\% to 3.4\%, which is 
comparable to the mass resolution of the electron channel, 3.9\% 
obtained without a kinematic fit. For the $Z\gamma$ 
mass range considered, photon energy contributions to the three-body mass 
resolution is much larger than that of the $Z$ boson width, which is
neglected in the kinematic fit.

Backgrounds to $Z\gamma$ production from the decay of a new particle 
include the SM $Z\gamma$ and
$Z$+jet processes, where the jet is misidentified as a photon. 
Backgrounds from processes with a photon where one or 
both of the leptons is due to a misidentified jet are found to be negligible. 
Contributions from 
$Z\gamma$ events with $Z\rightarrow \tau\tau$ and subsequent leptonic decays 
of the tau are less than 1\% of the sample. Contributions from 
$WZ$ and $ZZ$ processes, where electrons are misidentified as photons, are also
less than 1\% of the sample.

Efficiencies and background contributions are calculated 
using  independent data samples and Monte Carlo simulations. Scalar 
particle decays to $Z\gamma$ are modeled using 
{\sc pythia}~\cite{pythia} SM Higgs boson production in which the Higgs 
boson is 
forced to decay to $Z\gamma$, and the $Z$ boson is forced to decay into 
leptons.  
For the SM $Z\gamma$ events, we use an event generator employing 
first-order QCD calculations and first-order EW radiation~\cite{baur}. These
events are processed through a parameterized detector simulation that
is tuned on $Z$ boson candidate events. 
The background due to jets misidentified as 
photons is estimated by scaling the measured
$Z$+jet event rate by the measured probability for a jet to mimic a 
photon~\cite{us}.

\begin{figure}
\includegraphics[scale=0.43]{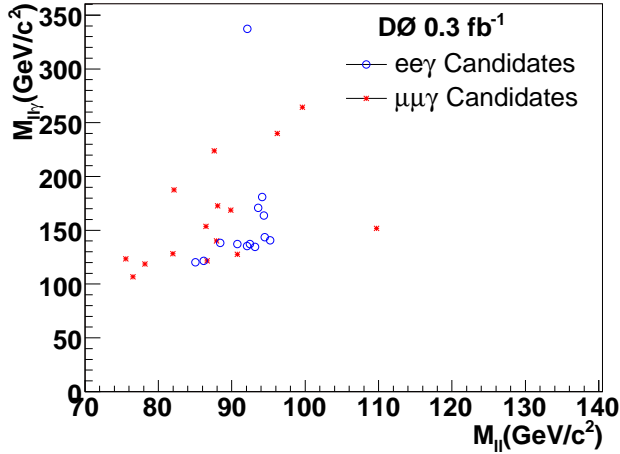}
\caption{
Distribution of candidates in the three-body mass, $M_{ll\gamma}$, vs
two-body mass, $M_{ll}$, plane is shown.  The electron candidates are blue circles and the muons are red starts.  The muon candidates are shown before the
two-body mass constraint is applied. 
\label{fig:mvm}}
\end{figure}

\begin{figure}
\includegraphics[scale=0.43]{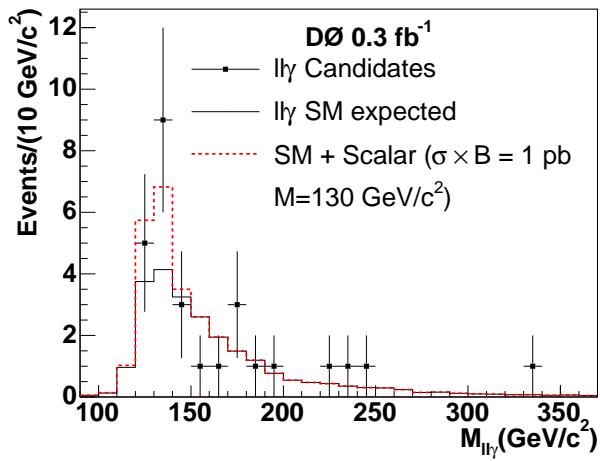}
\caption{
Distribution of the three-body mass, $M_{ll\gamma}$, for candidate
events and SM expectations. The signal shape for  a 130 GeV/$c^2$ 
scalar decaying to $Z\gamma$ with
a $\sigma(p \bar{p} \rightarrow X) \times B(X\rightarrow Z\gamma)$=1 pb
is also shown.
\label{fig:llg_dmch}}
\end{figure}

The final sample used in the analysis consists of 13 candidates in 
the electron channel and 15 candidates in the muon channel.  
We expect from SM sources 11.2 $\pm$ 0.8 
events in the electron channel and 12.9 $\pm$ 0.9 events
in the muon channel. 
Approximately 75\% of the expected SM contribution is due to SM $Z\gamma$.
Uncertainties in the SM contributions
are due to uncertainties in the luminosity, higher order QCD contributions, 
parton distribution functions, and the rate at which a jet
mimics a photon.  The luminosity uncertainty is the largest:
0.5 events for the electron channel
and 0.7 events for the muon channel.
In Fig~\ref{fig:mvm} we plot the three-body mass against the two-body
mass for the candidates.  The muon candidates are shown before the
two-body mass constraint is applied.  A single candidate fails the
$\chi^2$ cut for this constraint; it is the candidate with 
$M_{ll} =76$ GeV/$c^2$ and $M_{ll\gamma} = 107$ GeV/$c^2$. 
The $M_{ll\gamma}$ spectrum of the electron and muon data samples  
individually are consistent with the shapes of their respective Monte Carlo  
samples.  The three-body mass, $M_{ll\gamma}$, of the 
combined sample is shown in Fig.~\ref{fig:llg_dmch}.  The SM
expectations are also shown together with those for a 130 GeV/$c^2$ 
scalar decaying into $Z\gamma$ with $\sigma \times B$ of 1 pb.
This figure is just for illustration
purposes and is not used further in the analysis.


None of the 28 candidate events
 has more than one photon or more than two leptons.
 Among all the events we only find three jets with
$p_{T} >$15 GeV/$c$. 
Two of these jets are in a single event. The missing 
transverse momentum in all candidate events is less than 20 GeV/$c$.

We use two methods in our search to ensure sensitivity to scalar states over 
a broad range of natural decay widths.  The first looks for an excess in a 
sliding narrow window in the $M_{ll\gamma}$ spectrum, while the second sets a sliding 
lower mass threshold and counts events above this threshold. 
The window technique gives very good separation
of signal from background; however it is sensitive to the natural
width ($\Gamma$) of the new particle.  
The separation of signal from background of the window method is highest 
when $\Gamma$ is small compared to the mass resolution. 
The size of the search window was chosen to
be 4.4\% of the mass by optimization of the signal MC acceptance for a 
130 GeV/$c^2$ $Z\gamma$ resonance over the square-root of the SM background 
expectation.

The threshold technique also generally requires knowledge of 
$\Gamma$. To
reduce this dependence, we place the threshold at the median value of the
mass distribution ($M^\prime$), which introduces an acceptance factor of 0.5.  
The value of $M^\prime$ is the same as the nominal mass of the
 particle if its width
is fairly narrow ($\alt$  4 GeV/$c^2$), or if its mass 
is fairly low 
( $\alt$  250 GeV/$c^2$).  If neither condition is met, the 
available parton luminosity
begins to affect the generated mass distribution.   The SM Higgs boson 
provides a good example of this effect. A Higgs with a nominal mass of 
250 GeV/$c^2$ has a width of 4 GeV/$c^2$ and the median mass is 
249.7 GeV/$c^2$.
As the nominal mass increases, the width grows and the median mass begins to
deviate from the nominal value. 
At 350 and 450 GeV/$c^2$, the widths are 15 and 42 GeV/$c^2$, respectively; and the median masses are 
346.4 and 401.0 GeV/$c^2$, respectively.

Using these techniques, we determine the agreement between data and SM
expectations, taking into account systematic uncertainties.
Using the threshold technique, we find that the smallest probability of 
agreement between data and SM expectations is 7\%, which occurs at the 
median mass $M^\prime =$ 230 GeV/$c^2$.
Applying the narrow mass window method to search for objects with 
$\Gamma\rightarrow0$ (i.e. generated within the mass bin), 
we find that for a mass of 140 GeV/$c^2$, the 
probability of agreement between the data and SM expectation is 0.8\%.
The window at 140 GeV/$c^2$ has the lowest probability of agreement 
in the mass range considered.
To further assess the statistical significance of this effect, we generate 
an ensemble of 100,000 simulated experiments in which only SM sources for 
$Z\gamma$ were included and possible systematic effects are neglected.  
Eleven percent of the experiments contain a search window with a probability 
of agreement with the SM expectation of 0.5\% or less.  
The disagreement of 140 GeV/$c^2$ mass window has a significance of less than
$2.5$ standard deviations and lies at the mass where the SM 
background is largest and, therefore, where the ensemble tests indicate 
fluctuations would also be largest.

Since we find no excess 
in the data compared to the SM expectation, we extract limits on 
$\sigma(p\bar p \to X) \times B(X \to Z\gamma)$ for new scalar states.  
The limits are set using
a Bayesian technique~\cite{cox} with a flat prior for the signal and with 
systematic 
uncertainties on the signal and background taken into account.   

\begin{figure}
\includegraphics[scale=0.43]{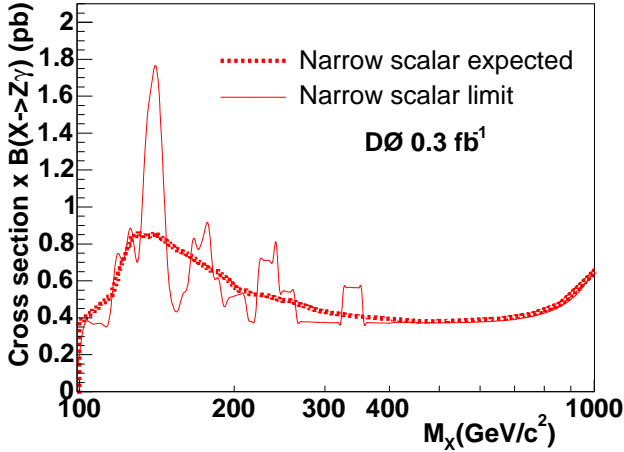}
\caption{
The expected and observed 
cross section times branching fraction 95\% C.L. limit 
for a scalar $X$ decaying into $Z\gamma$ as a function of $M$
for narrow scalar.
\label{fig:narexp}}
\end{figure}

\begin{figure}
\includegraphics[scale=0.43]{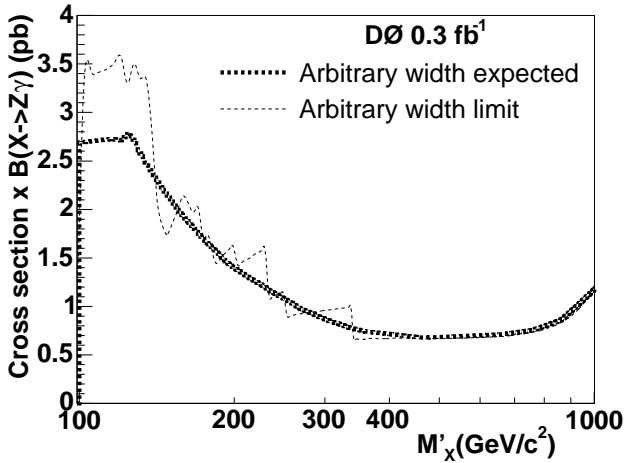}
\caption{
The expected and observed 
cross section times branching fraction 95\% C.L. limit 
for a scalar $X$ decaying into $Z\gamma$ as a function of $M^\prime$
for wide scalar.  $M^\prime$ is the median of the true mass distribution
for a generic object using the arbitrary width technique.
\label{fig:wideexp}}
\end{figure}

\begin{figure}
\includegraphics[scale=0.43]{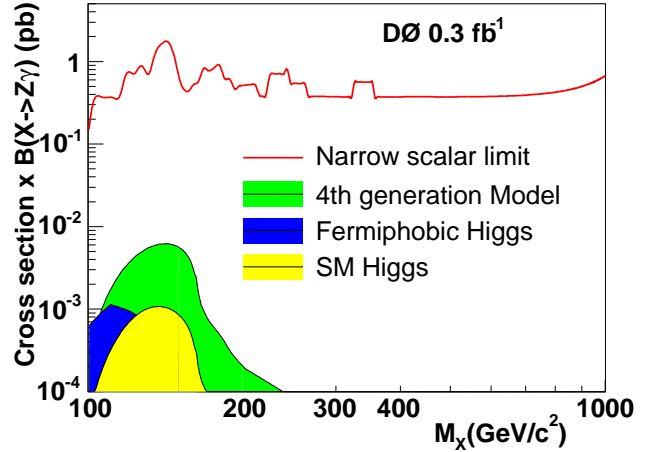}
\caption{
The cross section times branching fraction 95\% C.L. limits 
for a narrow scalar $X$ decaying into $Z\gamma$ as a function of M.
Curves representing the cross section times branching ratio expected from 
three variations of the Higgs are shown.   
\label{fig:alllim}}
\end{figure}

We extract the sensitivity and limits for two cases.
In the first
case, Fig.~\ref{fig:narexp},  we use the window technique 
and assume the width is negligible compared to the detector resolution.   
In the second
case, Fig.~\ref{fig:wideexp}, we use the threshold technique where 
the width is allowed to be at the other extreme. 
The 
expected limit for the window technique is less stringent where SM 
sources provide
the largest number of events; it is more stringent between 300 and 
800 GeV/$c^2$ where
no events are expected; it finally rises with mass as efficiency
decreases and the systematic uncertainties increase.
We see qualitatively similar structures
from the threshold technique limit. 
In comparing the two limits it should be noted that $M^\prime$ is
lower than the nominal mass of the particle.
  In Fig.~\ref{fig:alllim}, curves
representing the expected cross section times branching fraction for three
Higgs models are compared to the limits.  These models are the SM 
Higgs boson~\cite{spira}, a fermiophobic Higgs 
boson~\cite{Akeroyd:2005pr}, and  a model with four generations of 
quarks~\cite{Arik:2005ed}.  

In summary, we have performed the first search 
for $Z\gamma$ resonant states at a hadron collider
with an invariant mass greater than 100 GeV/$c^2$.  We find no statistically
significant evidence for the existence of these objects.  Narrowing our search
to scalar and pseudo-scalar resonances, we limit the production cross section
times branching fraction to less than 0.4 to 3.5 pb depending on the mass
and width.
%

%
We thank the staffs at Fermilab and collaborating institutions, 
and acknowledge support from the 
DOE and NSF (USA);
CEA and CNRS/IN2P3 (France);
FASI, Rosatom and RFBR (Russia);
CAPES, CNPq, FAPERJ, FAPESP and FUNDUNESP (Brazil);
DAE and DST (India);
Colciencias (Colombia);
CONACyT (Mexico);
KRF and KOSEF (Korea);
CONICET and UBACyT (Argentina);
FOM (The Netherlands);
PPARC (United Kingdom);
MSMT (Czech Republic);
CRC Program, CFI, NSERC and WestGrid Project (Canada);
BMBF and DFG (Germany);
SFI (Ireland);
The Swedish Research Council (Sweden);
Research Corporation;
Alexander von Humboldt Foundation;
and the Marie Curie Program.
%


\begin{thebibliography}{99}
%
\bibitem[*]{kurca}
On leave from IEP SAS Kosice, Slovakia.
\bibitem[\dag]{voutilainen}
Visitor from Helsinki Institute of Physics, Helsinki, Finland.
%
\vskip 0.25cm

\bibitem{Quigg:2005hr}  C.~Quigg,
  eConf {\bf C040802}, L001 (2004) [arXiv:hep-ph/0502070]; and 
references therein.

\bibitem{Feldman:2006ce}
  D.~Feldman, Z.~Liu and P.~Nath,
  arXiv:hep-ph/0603039.

  \bibitem{us} D\O\ Collaboration, V. M. Abazov {\it et al.}, Phys. Rev. Lett. {\bf 95}, 051802 (2005).
  \bibitem{otherD0} D\O\ Collaboration, B. Abbott {\it et al.}, Phys. Rev. D {\bf 57}, 3817 (1998); D\O\ Collaboration, S. Abachi {\it et al.}, Phys. Rev. Lett. {\bf 78}, 3640 (1997); D\O\ Collaboration, S. Abachi {\it et al.}, Phys. Rev. Lett. {\bf 75}, 1028 (1995).

  \bibitem{cdf} CDF II Collaboration,   D.~Acosta {\it et al.},
  Phys.\ Rev.\ Lett.\  {\bf 94}, 041803 (2005); CDF Collaboration, F. Abe {\it et al.}, Phys. Rev. Lett. {\bf 74}, 1941 (1995).

  \bibitem{delphi} DELPHI Collaboration, P. Abreu {\it et al.}, Phys. Lett. B {\bf 423}, 194 (1998).

  \bibitem{L3} L3 Collaboration,  P.~Achard {\it et al.},
Phys.\ Lett.\ B {\bf 597}, 119 (2004).

\bibitem{Abbiendi:2000cu}
  OPAL Collaboration, G.~Abbiendi {\it et al.}  ,
  Eur.\ Phys.\ J.\ C {\bf 17}, 553 (2000).

\bibitem{L3higgs} L3 Collaboration,  P.~Achard {\it et al.},
Phys.\ Lett.\ B {\bf 589}, 151 (2004).

\bibitem{spira} M. Spira, Report DESY T-95-05 (October 1995), arXiv:hep-ph/9510347.
\bibitem{Buscher:2005re}   V.~B\"uscher and K.~Jakobs,
  Int.\ J.\ Mod.\ Phys.\ A {\bf 20}, 2523 (2005).

\bibitem{Akeroyd:2005pr}
  A.~G.~Akeroyd, A.~Alves, M.~A.~Diaz and O.~J.~P.~Eboli,
  arXiv:hep-ph/0512077; A.~Alves, private communications for $\tan\beta=30$ 
and a charged Higgs mass of 100 GeV.
\bibitem{Arik:2005ed} E.~Arik, O.~Cakir, S.~A.~Cetin and S.~Sultansoy,
  Phys.\ Rev.\ D {\bf 66}, 033003 (2002).
\bibitem{Kozlov:2005rj}  G.~A.~Kozlov, 
  Phys.\ Rev.\ D {\bf 72}, 075015 (2005).
\bibitem{Ono:1983tf} S.~Ono, 
Acta Phys.\ Polon.\ B {\bf 15}, 201 (1984).
\bibitem{Cakir:2004nh}  O.~Cakir, R.~Ciftci, E.~Recepoglu, and S.~Sultansoy, 
  Acta Phys.\ Polon.\ B {\bf 35}, 2103 (2004).
%
\bibitem{Davoudiasl:2000wi}
   H.~Davoudiasl, J.~L.~Hewett, and T.~G.~Rizzo,
   Phys.\ Rev.\ D {\bf 63}, 075004 (2001)
%
\bibitem{run2det} V. Abazov {\it et al.}, accepted for publication by
Nucl. Instrum. Methods A, arXiv:physics/0507191.
\bibitem{coord} We use a cylindrical coordinate system about the beamline
in which positive $z$ is along the proton direction, $\theta$ is the polar
angle, $\phi$ is the azimuthal angle, and pseudorapidity ($\eta$) is defined
as $\eta=-\ln[\tan(\theta/2)]$. 

\bibitem{deltaR} In the D\O\ coordinate  system $\Delta {\cal{R}} =\sqrt{(\Delta\phi_{\ell\gamma})^2 +(\Delta\eta_{\ell\gamma})^2}$. 


\bibitem{pythia} T. Sj\"ostrand {\it et al.}, Computer Physics Commun. {\bf 135},
   238 (2001). Version 6.2.

\bibitem{baur} U. Baur and E. Berger, Phys. Rev. D {\bf 47}, 4889 (1993).

\bibitem{cox} R.T. Cox, Am. J. Phys. {\bf 14}, 1 (1946); H. Jeffreys, 
``Theory of Probability,'' 3rd edition, Oxford University Press, (1961).  


\end{thebibliography}
\end{document}